\documentclass[prb,twocolumn,showpacs,preprintnumbers,amsmath,amssymb,superscriptaddress]{revtex4}

\usepackage{graphicx}
\usepackage{dcolumn}
\usepackage{bm}
\def\Vec#1{\mbox{\boldmath $#1$}}

\begin{document}


\title{
Surface Bound States in $n$-band Systems with Quasiclassical Approach
}

\author{Yuki Nagai}
\affiliation{
Department of Physics, University of Tokyo, Tokyo 113-0033, Japan
}
\affiliation{JST, TRIP, Chiyoda, Tokyo, 102-0075, Japan}

\author{Nobuhiko Hayashi}
\affiliation{
Nanoscience and Nanotechnology Research Center (N2RC),
Osaka Prefecture University, 1-2 Gakuen-cho, Naka-ku, Sakai, 599-8570 Osaka, Japan
}
\affiliation{
CREST(JST), 4-1-8 Honcho, Kawaguchi, Saitama 332-0012, Japan
}

\date{\today}

\begin{abstract}
We discuss the tunneling spectroscopy  
at a surface
in multi-band systems such as Fe-based superconductors with the use of the quasiclassical approach. 
We extend the single-band method by 
Matsumoto and Shiba [J.\ Phys.\ Soc.\ Jpn.\ {\bf 64}, 1703 (1995)] into $n$-band systems ($n \ge 2$).
We show that the appearance condition of the zero-bias conductance peak
does not depend on details of the pair-potential anisotropy,
but it depends on details of the normal state properties 
in the case of fully-gapped superconductors.
The surface density of states in a two-band superconductor is presented as a simplest application.
The quasiclassical approach enables us to calculate readily the surface-angular dependence of the tunneling spectroscopy. 
\end{abstract}

\maketitle

\section{Introduction}
Much attention has been focused on novel Fe-based superconductors since the recent discovery
of superconductivity at the high temperature 26K in LaFeAsO$_{1-x}$F$_{x}$.\cite{Kamihara}
Many theoretical and experimental studies on Fe-based superconductors have been reported 
for the last year. 
It is important to identify the superconducting order parameter to elucidate the mechanism of 
superconductivity in those high-$T_{c}$ materials.

A $\pm s$-wave pairing symmetry has been theoretically proposed
as one of the candidates for the pairing symmetry in Fe-pnictide superconductors.\cite{bang0807,parish,kuroki,seo,Arita,mazin,eremin,nomura,stanev,senga} 
The $\pm s$-wave symmetry means that
the symmetry of pair potentials on each Fermi surface is $s$-wave
and the relative phase between them is $\pi$. 
Recently, we showed that a fully-gapped anisotropic $\pm s$-wave superconductivity 
consistently explained experimental observations such as nuclear magnetic relaxation rate and superfluid density. \cite{NagaiNJP}

A key point to identify the $\pm s$-wave symmetry is a detection of the sign change in 
the order parameters between Fermi surfaces. 
It is difficult to detect the relative phase of the order parameters in a bulk material.
However, as shown in studies of high-$T_{c}$ cuprates, Andreev bound states are formed at a surface or a junction when  
the quasiparticles feel different signs of the order parameter before and after scattering.\cite{TanakaPRL,Kashiwaya,Hu}
%
Since one can extract the information on the relative phase through Andreev bound states,
several theoretical studies on junctions and surfaces have been reported recently. \cite{Choi,Golubov,Araujo,onari,Wang,Linder,LinderPRB,Tsai,Ghaemi}
Andreev bound states at zero energy have been experimentally observed as a zero bias conductance peak (ZBCP) 
in tunneling spectroscopy for Fe-based superconductors.\cite{Yates}

The Fe-based superconductors are interesting also as novel unconventional multi-band superconductors 
since multi-band effects are essentially 
important there.\cite{kuroki}
Fermi surfaces in these systems predominantly consist of the $d$ orbitals of Fe atom. 
Kuroki {\it et al.}\cite{kuroki} suggest
that five orbitals are necessary to describe the properties of the superconductivity, and 
they elaborate an effective 5-band model. 
On the other hand, MgB$_{2}$ is a 2-band system that is a conventional $s$-wave BCS-like superconductor. 

The aim of this paper is to develop a method for analyzing surface bound states in multi-band superconductors.  
Matsumoto and Shiba\cite{Matsumoto} developed a method to analyze surface bound states in single-band systems 
such as high-$T_{c}$ cuprates. 
We extend their method into multi-band systems. 
Since the ratio of the superconducting gap $\Delta$ to the Fermi energy $E_{F}$ is small,
$\Delta/E_{F} \ll 1$, in 
Fe-based superconductors, 
we can adopt a quasiclassical approach.  
In this approach, all we need is only to consider quasiparticles at the Fermi level.
Thus, 
we can reduce 
computational machine-time and 
the physical picture becomes clear. %
In addition, this approach enables us to easily calculate 
the surface-angular dependence of 
tunneling spectroscopy. 
We find a general appearance condition of the ZBCP for multi-band systems. %
This general condition 
can be applied to 
various pairing symmetries including $\pm s$-wave and 
$d$-wave.
With our method, we will discuss 
a two-band superconductor as a simple example.

This paper is organized as follows. 
The formulation of our quasiclassical approach is shown in Sec.~II. 
We apply a quasiclassical approximation to 
eliminate fast spatial oscillations with Fermi wave length. 
The appearance condition of the ZBCP in multi-band systems is derived in Sec.~III. 
The results for a two-band model are shown as a simple example of our approach in Sec.~IV, where 
we will show both analytical and numerical results. 
The discussions and conclusion are given in Secs.~V and VI, respectively. 
In the appendix, we describe the derivation of the appearance
condition of the ZBCP when a system can be treated without
quasiclassical approximation.

\section{Formulation}
\subsection{Orbital representation and Band representation}
\begin{figure}
\includegraphics[width = 4cm]{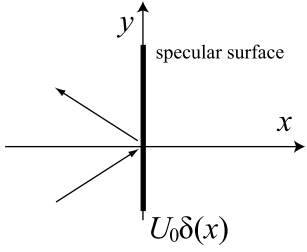}
\caption{\label{fig:spec}
Schematic figure of a specular surface.
}
\end{figure}
Let us consider the local density of states near a surface following
a procedure by Matsumoto and Shiba.\cite{Matsumoto}
We assume a two-dimensional superconductor,
and consider a specular surface,
for which the component of the quasiparticle momentum along the surface is conserved 
as shown in Fig.~\ref{fig:spec}. 
We treat the surface as a potential $U(\Vec{r}) \check{\tau}_{3}$, 
where the time-reversal symmetry is conserved.\cite{Matsumoto}
Here, $\check{\tau}_{i}$ ($i=1,2,3$) denote Pauli matrices in Nambu space
and $\Vec{r}$ is the position in the real space.
We consider a $n$-orbital system, which is a periodic crystal with $n$ atomic orbitals in unit cell.
Throughout the paper,
{\it hat} $\hat{a}$ denotes a $n \times n$ matrix in the orbital space, and {\it check} $\check{a}$ 
denotes a $2n \times 2n$ matrix composed of the $2 \times 2$ Nambu space
and the $n \times n$ orbital space. 
We calculate the Green function under the influence of $U(\Vec{r}) \check{\tau}_{3}$.
It is written as 
\begin{equation}
\check{G}(\Vec{r},\Vec{r}') = \check{G}_{0}(\Vec{r},\Vec{r}') + \int d\Vec{r}'' \check{G}_{0}(\Vec{r},\Vec{r}'') U(\Vec{r}'') \check{\tau}_{3} \check{G}(\Vec{r}'',\Vec{r}). \label{eq:green}
\end{equation}
Here, $\check{G}_{0}$ is an unperturbed Green function in the absence of $U$. 
We take the $x$($y$)-axis perpendicular (parallel) to the surface
as shown in Fig.~\ref{fig:spec}.  
Considering the surface situated at $x = 0$ and the scattering potential $U$ written as  $U(\Vec{r}) = U_{0} \delta(x)$, 
Eq.~(\ref{eq:green}) is reduced to 
\begin{equation}
\check{G}(x,k_{y},x',k'_{y}) = 2 \pi \delta(k_{y} - k'_{y}) \check{G}(x,x',k_{y}),
\end{equation}
where 
\begin{eqnarray}
\check{G}(x,x',k_{y}) &=& \check{G}_{0}(x,x', k_{y}) 
+ \check{G}_{0}(x,0,k_{y}) U_{0} \check{\tau}_{3} \nonumber \\
& & \times (1 - \check{G}_{0}(0,0,k_{y}) U_{0} \check{\tau}_{3}
)^{-1}
\check{G}_{0}(0,x',k_{y}).\nonumber \\
\end{eqnarray}
Here, we have taken the Fourier transformation with respect to $y$.  
We use units in which $\hbar = 1$, and the coordinates $\Vec{r}$ and the momentum $\Vec{k}$
are dimensionless. 
The surface is actually represented in the limit $U_{0} \rightarrow \infty$. 
The Green function is then given by 
\begin{equation}
\check{G}(x,x',k_{y}) = \check{G}_{0}(x,x', k_{y}) + \check{G}_{P}(x,x',k_{y}), \label{eq:4}
\end{equation}
where 
\begin{equation}
\check{G}_{P}(x,x',k_{y}) \equiv 
- \check{G}_{0}(x,0,k_{y})  \check{G}_{0}(0,0,k_{y})^{-1}
\check{G}_{0}(0,x',k_{y}). \label{eq:pg}
\end{equation}
The local density of states at the position $x$ for the momentum $k_{y}$ is written as 
\begin{equation}
N(x, k_{y}) = - \frac{1}{\pi} {\rm Im} \: [{\rm Tr} \:\hat{G}^{R}(x,x,k_{y})],
\label{eq:real-part-G}
\end{equation}
where 
\begin{equation}
\hat{G}^{R}(x,x,k_{y}) = \hat{G}(x,x,k_{y})|_{i \omega_{m} \rightarrow E + i \eta}.
\end{equation}
Here 
$\omega_{m}$ is the fermion Matsubara frequency and 
$\eta$ is a positive infinitesimal quantity. 
The unperturbed Green function $\check{G}_{0}^{R}(x,x',k_{y})$ is given by 
\begin{equation}
\check{G}_{0}^{R}(x,x',k_{y}) = \frac{1}{2 \pi} \int d k_{x} e^{i k_{x} (x- x')} \check{G}^{R}_{0}(k_{x},k_{y}), \label{eq:8}
\end{equation}
where
\begin{equation}
\check{G}_{0}^{R}(k_{x},k_{y}) = (E - \check{H}_{\rm N}^{o}(k_{x},k_{y}))^{-1}. \label{eq:greeno}
\end{equation}
Here, $ \check{H}_{\rm N}^{o}(k_{x},k_{y})$ is the $2n \times 2n$ Hamiltonian in Nambu and orbital spaces 
written as 
\begin{eqnarray}
\check{H}^{\rm o}_{\rm N} &\equiv& 
 \left(\begin{array}{cc}\hat{H}^o & \hat{\Delta}^o \\ \hat{\Delta}^{o \dagger} & - \hat{H}^o \end{array}\right), 
\end{eqnarray}
in the ``orbital representation'' where the base functions are atomic orbitals in crystal unit cell. 
From now on, the subscript ``$o$'' indicates that matrices are represented with the orbital basis. 
$\hat{H}^{o}$ is the Hamiltonian in the normal state represented
as $n \times n$ matrix in the orbital space.
Remember that $n$ is the number of the orbitals. 
$\hat{\Delta}^o$ is the superconducting order parameter.

Let us introduce a $n \times n$ Hamiltonian in the ``band representation'' defined by 
\begin{eqnarray}
\hat{H}^{b}(k_{x},k_{y}) &\equiv& \hat{P}^{-1}(k_{x},k_{y}) \hat{H}^{o}(k_{x},k_{y})\hat{P}(k_{x},k_{y}), \\
 &=& 
 \left(\begin{array}{ccc} \lambda_{1} & 0 & 0\\ 
 0 & \ddots & 0\\
0 &0 &\lambda_{n} \end{array}\right).
\end{eqnarray}
Here, $\lambda_{i}$ ($i=1,2,\cdots,n$) denote the eigenvalues where the relation $\lambda_{i} > \lambda_{j} \: \: (i < j)$ 
is satisfied. 
$\hat{P}$ is a unitary matrix consist of the eigenvectors that diagonalizes the Hamiltonian $\hat{H}^{o}$. 
The $2n \times 2n$ Hamiltonian in Nambu and orbital spaces in the ``band representation''  is also defined by 
\begin{eqnarray}
\check{H}_{\rm N}^{b}(k_{x},k_{y}) &\equiv& 
\check{U}^{-1}(k_{x},k_{y}) \check{H}_{\rm N}^{o}(k_{x},k_{y}) \check{U}(k_{x},k_{y}), \label{eq:bhami}\\
&=&  \left(\begin{array}{cc}\hat{H}^b & \hat{\Delta}^b \\ \hat{\Delta}^{b \dagger} & - \hat{H}^b \end{array}\right),
\end{eqnarray}
where 
\begin{eqnarray}
\check{U}(k_{x},k_{y}) &\equiv&  
\left(\begin{array}{cc}\hat{P}(k_{x},k_{y}) &0 \\ 
0& \hat{P}(k_{x},k_{y}) \end{array}\right), \\
\hat{\Delta}^{b} &\equiv& \hat{P}^{-1} \hat{\Delta}^{o} \hat{P}.
\end{eqnarray}
In general, $\hat{\Delta}^{b}$ contains off-diagonal elements, which correspond to inter-band pairings. 
Assuming that intra-band pairings are dominant, we neglect the off-diagonal (inter-band) elements in $\hat{\Delta}^{b}$:
\begin{eqnarray}
\hat{\Delta}^{b} &\approx& 
 \left(\begin{array}{ccc} \Delta_{1} & 0 & 0\\ 
 0 & \ddots & 0\\
0 &0 &\Delta_{n} \end{array}\right). \label{eq:dede}
\end{eqnarray}
That is, we consider that only single pair-potential is defined on each Fermi surface.
Here, $\Delta_{i}$ is the pair-potential on the $i$-th band. 
Substituting Eq.~(\ref{eq:bhami}) into Eq.~(\ref{eq:greeno}), 
the Green function $\check{G}_{0}^{R}(k_{x},k_{y})$ is written as 
\begin{equation}
\check{G}_{0}^{R}(k_{x},k_{y}) = \check{U} (E - \check{H}^{b})^{-1} \check{U}^{-1}.
\end{equation}
Assuming Eq.~(\ref{eq:dede}) and taking the inverse matrix of $E - \check{H}^{b}$, 
one can obtain
\begin{eqnarray}
\check{G}_{0}^{R}(k_{x},k_{y}) =  \check{U} 
\left(\begin{array}{cc}\hat{A}_{+} &\hat{B} \\ 
\hat{B}^{\dagger}& \hat{A}_{-}\end{array}\right)
 \check{U}^{-1}, \label{eq:ugu}
\end{eqnarray}
where 
\begin{eqnarray}
\hat{A}_{\pm} &=& 
 \left(\begin{array}{ccc} \frac{E \pm \lambda_{1}}{-|\Delta_{1}|^{2} +E^{2} - \lambda_{1}^{2}} & 0 & 0\\ 
 0 & \ddots & 0\\
0 &0 &\frac{E \pm \lambda_{n}}{-|\Delta_{n}|^{2} +E^{2} - \lambda_{n}^{2}} 
\end{array}\right), \\
\hat{B} &=& 
 \left(\begin{array}{ccc} \frac{\Delta_{1}}{-|\Delta_{1}|^{2} +E^{2} - \lambda_{1}^{2}} & 0 & 0\\ 
 0 & \ddots & 0\\
0 &0 &\frac{\Delta_{n}}{-|\Delta_{n}|^{2} +E^{2} - \lambda_{n}^{2}} 
\end{array}\right).
\end{eqnarray}
We find that Eq.~(\ref{eq:ugu}) can be rewritten as 
\begin{eqnarray}
\check{G}_{0}^{R}(k_{x},k_{y}) &=& \sum_{i} \check{G}^{i}(k_{x},k_{y}),  \label{eq:22}
\end{eqnarray}
where $i$ is the band index and
\begin{eqnarray}
\check{G}^{i}&\equiv&
\frac{
1
}{-|\Delta_{i}|^{2} + E^{2} - \lambda_{i}^{2}} 
\left(\begin{array}{cc}(E + \lambda_{i}) \hat{M}_{i} &\Delta_{i} \hat{M}_{i} \\ 
\Delta^{\ast}_{i} \hat{M}_{i}& ( E - \lambda_{i}) \hat{M}_{i} \end{array}\right),  \nonumber \\
\label{eq:23}
\\
{[} \hat{M}_{i} {]}_{jk} &=& {[} \hat{P} {]}_{ji} {[} \hat{P} {]}_{ki}^{\ast }. 
\end{eqnarray}
Equation (\ref{eq:22}) is divided into a sum of the Green functions defined on each band. 
Substituting Eq.~(\ref{eq:22}) into Eq.~(\ref{eq:8}), $\check{G}_{0}^{R}(x,x',k_{y})$ is expressed as 
\begin{equation}
\check{G}_{0}^{R}(x,x',k_{y}) = \sum_{i} \frac{1}{2 \pi} \int d k_{x} e^{i k_{x} (x- x')} \check{G}^{i}(k_{x},k_{y}). \label{eq:g0wa}
\end{equation}
Hence, the $k_{x}$-integration is found to be performed on each band independently.
%
\subsection{Quasiclassical Approach}
We assume $|\Delta_{i}| \ll E_{F}$. 
This relation is satisfied in 
most of systems such as conventional superconductors and Fe-based ones. 
In this case, one can use a quasiclassical approach. 

\begin{figure}
\includegraphics[width = 8cm]{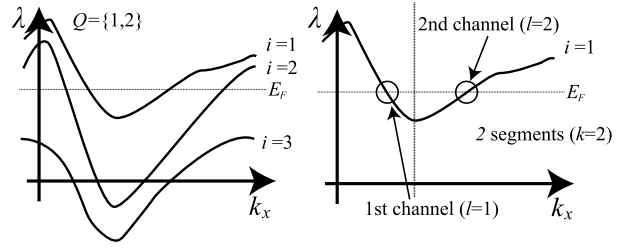}
\caption{\label{fig:segments}
Schematic figures of band dispersions along a $k_{x}$ line with a fixed $k_{y}$.
}
\end{figure}

We consider a line with a fixed $k_{y}$ in the momentum space.   
On this line, we classify $n$-bands into two groups. 
One group is composed of the bands on 
which the eigen energy $\lambda_{i}(k_{x},k_{y})$ %
crosses the Fermi level (for example, the bands $i = 1$ and 2 in Fig.~\ref{fig:segments}).
The other group is composed of the bands on which the eigen energy does not cross the Fermi level
(the band $ i = 3$).
For the former group, we can analytically integrate $\check{G}^{i}(k_{x},k_{y})$ over $k_{x}$ with the use of a quasiclassical approach
since $\check{G}^{i}(k_{x},k_{y})$ is a function localized near the Fermi level.
For the latter group, we need to integrate $\check{G}^{i}(k_{x},k_{y})$ over $k_{x}$ numerically
since $\check{G}^{i}(k_{x},k_{y})$ is not a localized function. 
However, the integrand is a smooth function, so that 
it is easy to perform such a numerical integration.  

We integrate $\check{G}^{i}(k_{x},k_{y})$ on the bands of the first group with the use of the quasiclassical approach.
To perform the $k_{x}$-integration, we divide the $k_{x}$-line with a fixed $k_{y}$ into some segments 
as shown in Fig.~\ref{fig:segments}. %
Each segment has only single channel that is the point satisfying the relation $\lambda_{i} = E_{F}$. 
From now on, 
$l$ denotes the channel index and $k$ denotes the maximum number of $l$.  
The integration for the $i$-th band is written as 
\begin{equation}
\int dk_{x} \sim \sum_{l=1}^{k} \int_{- \infty}^{\infty} \frac{d \lambda_{i}}{v_{i}(\lambda_{i})}.
\end{equation}
Expanding $k_{x}(\lambda_{i})$ in the first order of $\lambda_{i}$
around $\lambda_{i} = E_{F}$ as $k_x(\lambda_i) = k_{Fx} + \lambda_i/v_{Fx}$,
one can carry out the integration by the residue theorem:
\begin{equation}
\frac{1}{2 \pi} \int d k_{x} e^{i k_{x} (x- x')} \check{G}^{i}(k_{x},k_{y}) = - i \sum_{l=1}^{k} \check{G}^{F}_{i,l}(k_{Fx}^{i,l}),
\end{equation}
where
\begin{eqnarray}
\check{G}^{F}_{i,l}(k_{Fx}^{i,l}) &\equiv&
\frac{e^{i k_{Fx}^{i,l}(x-x')}e^{i |x-x'|\frac{\sqrt{E^{2} - |\Delta_{i}|^{2}}}{|v_{Fx}^{i,l}|}}
}{2 |v_{Fx}^{i,l}| \sqrt{E^{2} - |\Delta_{i}|^{2}}} \check{F}(k_{Fx}^{i,l}), \label{eq:gf}\\
\check{F}(k_{Fx}^{i,l}) &\equiv&  \left(\begin{array}{cc} 
f_{+}(k_{Fx}^{i,l})\hat{M}_{i}(k_{Fx}^{i,l})  & \Delta_{i} (k_{Fx}^{l}) \hat{M}_{i}(k_{Fx}^{i,l})  \\ 
\Delta_{i} (k_{Fx}^{i,l}) \hat{M}_{i}(k_{Fx}^{i.l})  &f_{-}(k_{Fx}^{i,l})\hat{M}_{i}(k_{Fx}^{i,l}) 
 \end{array}\right), \nonumber \\ \\
 f_{\pm}(k_{Fx}^{i,l}) &\equiv& 
 E \pm {\rm sgn} \: (x-x') {\rm sgn} \: (v_{Fx}^{i,l}) \sqrt{E^{2} - |\Delta_{i}|^{2}}, \nonumber \\
\end{eqnarray}
Here, %
$k_{Fx}^{l}$ and $v_{Fx}^{l}$ are
the Fermi wave number
and the Fermi velocity on the $l$-th channel, respectively.
Using the above, Eq.~(\ref{eq:g0wa}) can be written as 
\begin{eqnarray}
\check{G}_{0}^{R}(x,x',k_{y}) &=& -i \sum_{i \in Q}  \sum_{l=1}^{k} \check{G}^{F}_{i,l}(x,x',k_{Fx}^{i,l}) \nonumber \\ 
& & + 
\sum_{i  \notin Q} \frac{1}{2 \pi} \int d k_{x} e^{i k_{x} (x- x')} \check{G}^{i}(k_{x},k_{y}), \nonumber \\ \label{eq:g0quasi}
\end{eqnarray}
where the elements in $Q$ are the indices of the bands whose energy dispersions cross the Fermi level for a fixed $k_y$.  
Here, we assume $\Delta_{i \notin Q} = 0$, namely the superconducting order parameters are finite
only around the Fermi level. 
It should be noted that the second term in the right-hand side of Eq.~(\ref{eq:g0quasi}) cannot be neglected since 
$\check{G}^{R}_{0}(0,0,k_{y})^{-1}$ without this second term may have artificial divergences. 
\subsection{Eliminating the fast oscillations with Fermi wave length}
We assume the condition $k_{F} \xi \gg 1$ (i.e., $|\Delta_{i}| \ll E_{F}$), which is the quasiclassical condition. 
Here, $\xi$ is the coherence length of a superconductor.
Under this condition, the short range spatial oscillations characterized by the Fermi wave length $1/k_{F}$ can 
be eliminated. 
We rewrite Eq.~(\ref{eq:g0quasi}) as 
\begin{equation}
\check{G}_{0}^{R}(x,x',k_{y}) = \sum_{i} \int dk_{x} \check{K}_{i}(k_{x},k_{y}) e^{i k_{x} (x-x')}, 
\end{equation}
where
\begin{eqnarray}
\check{K}_{i \in Q}(k_{x},k_{y}) &\equiv& -i \sum_{l}^{k} G^{F}_{i,l}(x,x',k_{x}) \delta(k_{x} - k_{Fx}^{i,l}) , \: \: \: \; \; \; \\
\check{K}_{i \notin Q}(k_{x},k_{y}) &\equiv&   \frac{1}{2 \pi}   \check{G}^{i}(k_{x},k_{y}).
\end{eqnarray}
The perturbed Green function $\check{G}_{P}(x,x',k_{y})$ defined in Eq.~(\ref{eq:pg}) can be written as 
\begin{eqnarray}
\check{G}_{P}^{R}(x,x',k_{y}) &=& - \sum_{i,i''} \int dk_{x} d k_{x}''  e^{i (k_{x} x - k_{x}'' x')} 
 \check{K}_{i}(k_{x},k_{y}) \nonumber \\
 & & \times 
\check{G}_{0}^{R}(0,0,k_{y})^{-1} 
  \check{K}_{i''}(k_{x}'',k_{y}) .
\end{eqnarray}
Setting $\exp[i (k_{x} x - k_{x}'' x')] \rightarrow 1$, we eliminate the short range oscillation
while keeping the enveloping profile of the integrand. 
Thus, the above equation is reduced to 
\begin{eqnarray}
\check{G}_{P}^{R}(x,x',k_{y}) &=& - \sum_{i} \int dk_{x} 
 \check{K}_{i}(k_{x},k_{y}) 
\check{G}_{0}^{R}(0,0,k_{y})^{-1} \nonumber \\
& & \times  
  \sum_{i''} \int dk_{x}'' \check{K}_{i''}(k_{x}'',k_{y}).
\label{eq:36}
\end{eqnarray}
From this equation, it is concluded that the Andreev bound states appear when $\check{G}_{0}^{R}(0,0,k_{y})^{-1}$ diverges,
i.e., when $\det \check{G}_{0}^{R}(0,0,k_{y}) =0$.
\section{Appearance condition of the Zero Bias Conductance Peak (ZBCP)}
Let us consider the appearance condition of the ZBCP in $n$-band system at a surface. 
At the zero energy $E=0$, $\check{G}^{F}_{i,l}(x=0,x'=0,k_{Fx}^{l,i})$ defined in Eq.~(\ref{eq:gf}) [for $i \in Q$] is written as 
\begin{eqnarray}
\check{G}^{F}_{i,l}(k_{Fx}^{i,l}) &=&
\frac{{\rm sgn} \: (\Delta_{i})}
{2 |v_{Fx}^{i,l}| }
\left(\begin{array}{cc} 
0  & \hat{M}_{i}(k_{Fx}^{i,l})  \\ 
\hat{M}_{i}(k_{Fx}^{i.l})  & 0
 \end{array}\right).
\end{eqnarray}
For $i \notin Q$, we have from Eq.\ (\ref{eq:23}) with $E=0$,
\begin{eqnarray}
\check{G}^{i}
=
\frac{1}{ - \lambda_{i}^{2}} 
\left(\begin{array}{cc}\lambda_{i} \hat{M}_{i} & 0 \\ 
0 & - \lambda_{i} \hat{M}_{i} \end{array}\right),
\end{eqnarray}
where we have set $\Delta_{i}=0$ because the superconducting order parameter is assumed to be finite
only near the Fermi level and the bands with the indices $i \notin Q$ do not cross it.
Substituting the above equations into Eq.~(\ref{eq:g0quasi}), we can obtain the appearance condition of 
the ZBCP from $\det \check{G}_{0}^{R}(0,0,k_{y}) =0$: 
\begin{eqnarray}
{\rm det}\: 
\left(\begin{array}{cc} 
-\hat{I}  & \hat{L} \\ 
\hat{L} & \hat{I}
 \end{array}\right)
 = 0, \label{eq:app}
\end{eqnarray}
where
\begin{eqnarray}
\hat{L} &\equiv& -i \sum_{i \in Q} \sum_{l} \frac{{\rm sgn} \: (\Delta_{i}(k_{Fx}^{i,l}))}
{2 |v_{Fx}^{i,l}| }\hat{M}_{i}(k_{Fx}^{i,l}), \label{eq:l}\\
\hat{I} &\equiv& \sum_{i \notin Q} \frac{1}{2 \pi} \int \frac{d k_{x}}{\lambda_{i}(k_{x})} \hat{M}_{i}(k_{x}). \label{eq:hat-I}
\end{eqnarray}
Equation (\ref{eq:l}) shows that the appearance condition {\it does not} depend on the anisotropy of the pair-potentials and it depends only on the signs of them 
because information on the pair potentials is included
in the form, ${\rm sgn} \: (\Delta_{i}(k_{Fx}^{i,l}))$, in Eq.~(\ref{eq:l}). 
This result shows that information on the normal state (i.e.,  the matrices $\hat{M}_{i}$, $v_{Fx}^{i,l}$) is important 
for the ZBCP to appear.
\section{Two-band model as a simple example}
\subsection{Model}
\begin{figure}
  \begin{center}
    \begin{tabular}{p{40mm}p{40mm}}
      \resizebox{55mm}{!}{(a)\includegraphics{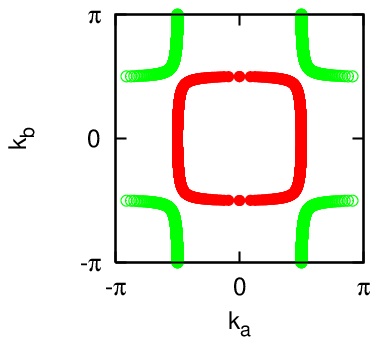}} &
      \resizebox{55mm}{!}{(b)\includegraphics{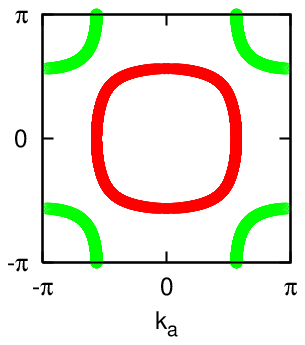}} 
    \end{tabular}
\caption{\label{fig:fer}
(Color online)
Fermi surfaces in the two-band model.
(a) the half filling ($\mu = 0$) and $t' = 0.1 t$.
(b) $\mu = 0.2t$ and $t' = 0.2t$.
}
  \end{center}
\end{figure}

We calculate the density of states in a two-band superconductor as a simple example.
We consider a two-band tight-binding model on a square lattice.
There are two orbitals on each lattice site.
The Hamiltonian with a $2\times2$ matrix form in the normal state is described as 
\begin{eqnarray}
\hat{H}^{o} = 
\left(\begin{array}{cc} 
-t \cos(k_{a}) - \mu & 2 t' \sin(k_{a}) \sin (k_{b}) \\ 
2 t' \sin(k_{a}) \sin(k_{b}) & - t \cos(k_{b}) -\mu
 \end{array}\right),
\end{eqnarray}
in the orbital representation ($n=2$).
Here, $k_{a}$ and $k_{b}$ are the axes fixed to the crystal axes in the momentum space, $t$ and $t'$ are 
intra- and inter-orbital hopping amplitudes, respectively, 
and $\mu$ denotes the chemical potential.
We use the unit in which the lattice constant $a = 1$.  
This Hamiltonian can be diagonalized into the matrix in the band representation, $\hat{H}^{b}$, written as 
\begin{eqnarray}
\hat{H}^{b} = \hat{P}^{-1} \hat{H}^{o} \hat{P} = 
\left(\begin{array}{cc} 
\lambda_{A} & 0 \\ 
0 & \lambda_{B}
 \end{array}\right).
\end{eqnarray}
Here, $\lambda_{A(B)}$ denotes the energy dispersion on the $A(B)$-band.
As shown in Fig.~\ref{fig:fer}, the Fermi surfaces consist of two parts near the half filling.

We consider the two-band $s$-wave superconductor described by the pair potential in the band representation: 
\begin{eqnarray}
\hat{\Delta}^{b} = 
\left(\begin{array}{cc} 
\Delta_{A}& 0 \\ 
0 & \Delta_{B}
 \end{array}\right).
\end{eqnarray}
Here, $\Delta_{A(B)}$ is the pair potential on the $A(B)$-band.

We introduce the coordinates ($k_{a}$,$k_{b}$) fixed to the crystal axes: 
\begin{eqnarray}
\left(\begin{array}{c} 
k_{a}  \\ 
k_{b} 
 \end{array}\right) = 
 \left(\begin{array}{cc} 
\cos \theta & - \sin \theta \\ 
\sin \theta & \cos \theta
 \end{array}\right)
 \left(\begin{array}{c} 
k_{x}  \\ 
k_{y} 
 \end{array}\right).
\end{eqnarray}
Here, the $k_{x}$($k_{y}$) axis is the axis parallel (perpendicular) to the surface
and 
 $\theta$ is the angle between the $k_{a}$ and $k_{x}$ axes.
Considering [110] surface, we fix $\theta = \pi/4$. 
The quasiparticle momentum $k_{y}$ is conserved since we consider the specular surface. 

It should be noted that one needs to treat the Brillouin zone in the surface-coordinates
$(k_{x},k_{y})$
for each surface angle
since it is necessary to consider all possible scattering processes at the specular surface (namely, all $k_y$-momentum conserving processes).
For example, it naively seems in Fig.~3(a) that possible scattering processes occur
only on the inner Fermi surface (red) for the [110] surface ($\theta = \pi/4$) in the region $\pi \sqrt{2}/4  < k_y <  \pi \sqrt{2}/2$ [the $k_y$ axis is directed in the direction of $(k_a,k_b)=(-1,1)$ in Fig.~3(a)].
However, for [110] surface, one has to consider also the outside of the first Brillouin zone as shown in Fig.~4 so that scattering process between outer Fermi surface (green) and inner Fermi surface (red) can occur.

\begin{figure}
\includegraphics[width = 5cm]{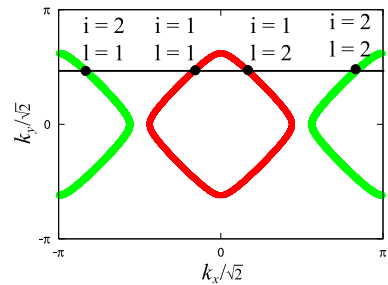}
\caption{\label{fig:yokogiru}
(Color online)
Fermi surfaces ($\mu = 0$ and $t' = 0.1 t$) and $k_x$ line with fixed $k_{y}$.
}
\end{figure}

At the half filling for [110] surface, the second term in Eq.~(\ref{eq:g0quasi}) does not exist 
since the energy dispersions of the A and B bands always cross the Fermi level on $k_x$ line with any fixed $k_{y}$ in the momentum space 
as shown in Fig.~\ref{fig:yokogiru}.
In this case, ${\hat I}$ defined in Eq.\ (\ref{eq:hat-I}) is zero
because there is no band with the index $i \notin Q$.
Therefore, the appearance condition of the ZBCP in Eq.~(\ref{eq:app}) can be rewritten as 
\begin{equation}
{\rm det} \: \hat{L} = 0 \label{eq:apptwo}.
\end{equation}
where $\hat{L}$ is defined in Eq.~(\ref{eq:l}). 

\subsection{Analytical Results}
\subsubsection{At the half filling for {\rm [110]} surface}
We will analytically show that the ZBCP always appears for any strength of the inter-orbital hopping $t'$ in the case of [110] surface ($\theta = \pi/4$) at the half filling.
On the lines which satisfy $k_a = (k_x-k_y)/\sqrt{2} = n \pi$ or $k_b = (k_x+k_y)/\sqrt{2} = n \pi$ in the momentum space, 
one can easily obtain the unitary matrix $\hat{P}$ that diagonalizes $\hat{H}^{o}$: 
\begin{eqnarray}
\hat{P}(k_y) &=& 
\left\{
\begin{array}{l}
 \left(\begin{array}{cc} 
1 & 0 \\ 
0 & 1
 \end{array}\right),  k_x - k_y = n_e \pi,  \: {\rm or} \: k_x + k_y = n_o \pi, \\ \nonumber \\ 
 \left(\begin{array}{cc} 
0 & 1 \\ 
1 & 0
 \end{array}\right), k_x - k_y = n_o \pi,  \: {\rm or} \: k_x + k_y = n_e \pi,
\end{array} 
\right. \label{eq:46}\\
\end{eqnarray}
Here, $n_{e(o)}$ is an even (odd) integer. 
Substituting these $\hat{P}(k_{y})$ into Eq.~(\ref{eq:l}), 
we obtain $\hat{L}$: 
\begin{eqnarray}
\hat{L} \propto \frac{{\rm sgn} \: (\Delta_{A}) + {\rm sgn} \: (\Delta_{B})}{|v_{Fx}|}
 \left(\begin{array}{cc} 
1 & 0 \\ 
0 & 1
 \end{array}\right).
\end{eqnarray}
The appearance condition of the ZBCP [Eq.~(\ref{eq:apptwo})] is written as 
\begin{equation}
{\rm sgn} \: (\Delta_{A}) + {\rm sgn} \: (\Delta_{B}) = 0, \label{eq:anare}
\end{equation}
on the lines where $k_a = (k_x-k_y)/\sqrt{2} = n \pi$ or $k_b = (k_x+k_y)/\sqrt{2} = n \pi$. 
This condition is always satisfied in the sign-reversing $s$-wave ($\pm s$-wave) superconductors in this model. 
The $\pm s$-wave symmetry means that the symmetry of pair potentials on each Fermi surface is
s-wave and the relative phase between them is $\pi$.\cite{bang0807,parish,kuroki,seo,Arita,mazin,eremin,nomura,stanev,senga,NagaiNJP}
Therefore, the ZBCP appears at the points on the Fermi surfaces
where the relation $k_{a} = 0$ or $k_{b} = 0$ is satisfied in the momentum space.

\subsubsection{Case of $t'/t = 0$ for {\rm [110]} surface}
In the case of $t'/t = 0$, we can analytically show that the ZBCP always occurs for [110] surface. 
In this case, the unitary matrix $\hat{P}$ can be written as 
\begin{eqnarray}
\hat{P}(k_y) &=& 
\left\{
\begin{array}{l}
 \left(\begin{array}{cc} 
1 & 0 \\ 
0 & 1
 \end{array}\right), \: \: \: \: k_x  k_{y} > 0, \\ \label{eq:uni}\\
 \left(\begin{array}{cc} 
0 & 1 \\ 
1 & 0
 \end{array}\right), \: \: \: \: k_x k_{y} < 0. 
\end{array} 
\right. 
\end{eqnarray}
As in the case of Eq.~(\ref{eq:46}), these unitary matrices lead to 
the same appearance condition of the ZBCP as Eq.~(\ref{eq:anare}).

\subsubsection{Case of $t'/t = 0$ at the half filling for {\rm [110]} surface}
Finally, we discuss the difference between the appearance conditions with and without the quasiclassical approach. 
As shown in Appendix, the appearance condition obtained without the quasiclassical approach
for $t' = 0$ at the half filling
is written as 
\begin{eqnarray}
\Delta_{ab} &=& 0, \label{eq:dab}\\
I_{1} = 0 \: \: &{\rm or}& \: \: I_{2} = 0,\label{eq:iab}
\end{eqnarray}
where
\begin{eqnarray}
I_{1,2} &=& 
\frac{\ln \left( 
\frac{ (\sin (\pm  k_{y}/\sqrt{2})  +\sqrt{1 + |\Delta_{A}/t|^{2}})^{2}}{(\sin (\pm  k_{y}/\sqrt{2})  -\sqrt{1 + |\Delta_{A}/t|^{2}})^{2}}
\right)}{ 2\sqrt{1 + |\Delta_{A}/t|^{2}}}    \nonumber \\
& & - \frac{ \ln \left( 
\frac{ (\sin (\pm  k_{y}/\sqrt{2})  +\sqrt{1 + |\Delta_{B}/t|^{2}})^{2}}{(\sin (\pm  k_{y}/\sqrt{2})  -\sqrt{1 + |\Delta_{B}/t|^{2}})^{2}}
\right)}{ 2\sqrt{1 + |\Delta_{B}/t|^{2}}},\label{eq:qui}\\
\Delta_{ab} &=& -\pi \left ( \frac{ {\rm sgn} \: (\Delta_{B}/t) 
 } 
 { \sqrt{ 1  +|\Delta_{A}/t|^{2}   }}
  + \frac{ {\rm sgn} \: (\Delta_{B}/t) } { \sqrt{ 1  +|\Delta_{B}/t|^{2} }  } \right). \label{eq:qud}
\end{eqnarray}
Here, we assume that the pair-potentials $\Delta_{A}$ and $\Delta_{B}$ do not depend on $\Vec{k}$ for simplicity. 
The above equations suggest that the appearance condition of the ZBCP depends on
the details of the amplitudes $|\Delta_{A}|$ and 
$|\Delta_{B}|$
in contrast to the quasiclassical result [Eq.~(\ref{eq:anare})].
In the limit of $|\Delta_{A,B}/t| \ll 1$, on the other hand, Eqs.~(\ref{eq:qui}) and (\ref{eq:qud}) are reduced to Eq.~(\ref{eq:anare}) obtained by the quasiclassical approximation. 
Thus, the quasiclassical and non-quasiclassical results coincide in this limit.
Therefore, it is suggested that our quasiclassical approach is appropriate when $|\Delta_{A,B}|/t \ll 1$. 

\subsection{Numerical Results}
The density of states at the surface is
calculated from Eq.~(\ref{eq:real-part-G})
as 
\begin{equation}
N(E) = \frac{1}{2 \pi} \int dk_{y} N(x = 0,k_{y}).
\end{equation}
We consider the $\pm s$-wave superconductor\cite{bang0807,parish,kuroki,seo,Arita,mazin,eremin,nomura,stanev,senga,NagaiNJP}
and the same two-band model as discussed in this section.

\subsubsection{Dependence of the surface-angle $\theta$}
\begin{figure}
  \begin{center}
    \begin{tabular}{p{40mm}p{40mm}}
      \resizebox{50mm}{!}{\includegraphics{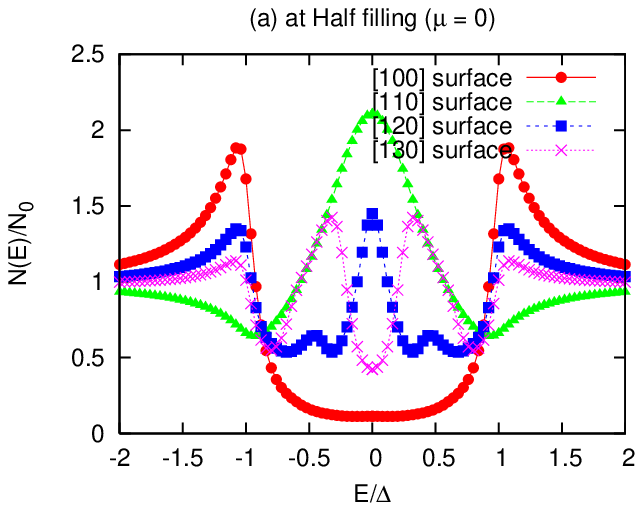}} &
      \resizebox{50mm}{!}{\includegraphics{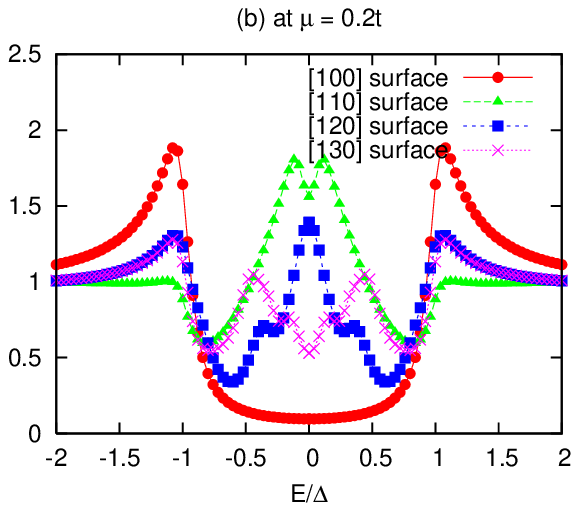}} 
    \end{tabular}
\caption{\label{fig:surdep}
(Color online)
The density of states at the surface for various surface angles.
The pair potentials are  $\Delta_{A}  = \Delta = 0.001t$ and $\Delta_{B} = -\Delta_{A}$.
(a) the half filling ($\mu = 0$) and (b) $\mu = 0.2t$.  
The inter-orbital hopping amplitude is $t' = 0.1t$.
The smearing factor is $\eta=0.1\Delta$.
}
  \end{center}
\end{figure}

\begin{figure}
  \begin{center}
    \begin{tabular}{p{40mm}p{40mm}}
      \resizebox{50mm}{!}{\includegraphics{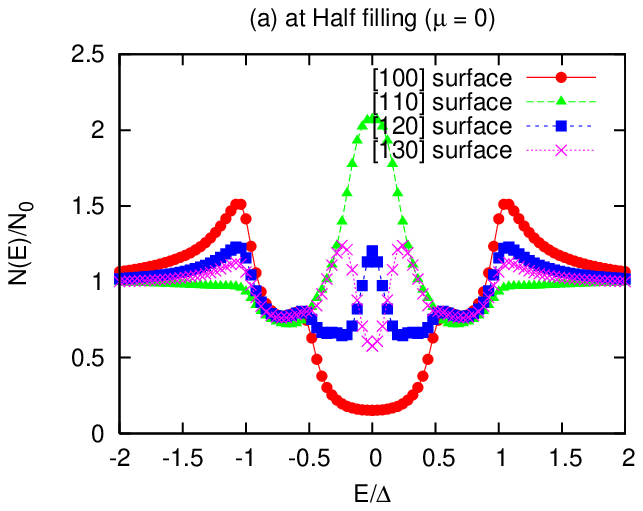}} &
      \resizebox{50mm}{!}{\includegraphics{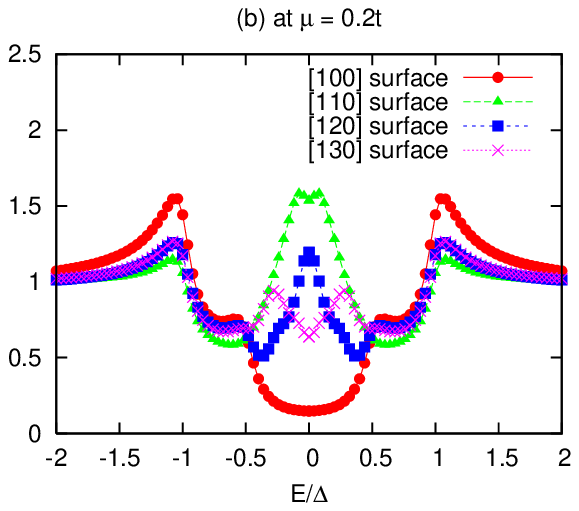}} 
    \end{tabular}
\caption{\label{fig:surdep2}
(Color online)
The density of states at the surface for various surface angles.
The pair potentials are  $\Delta_{A}  = \Delta = 0.001t$ and $\Delta_{B} = -0.5 \Delta_{A}$.
(a) the half filling ($\mu = 0$) and (b) $\mu = 0.2t$.  
The inter-orbital hopping amplitude is $t' = 0.1t$.
The smearing factor is $\eta=0.1\Delta$.
}
  \end{center}
\end{figure}

We show the energy dependence of the density of states for various surface-angle $\theta$ in Figs.~\ref{fig:surdep} 
and \ref{fig:surdep2}.
The peak positions of the Andreev bound states depend on the surface angle $\theta$.
By comparing the results between Figs.~\ref{fig:surdep} and \ref{fig:surdep2},
it is noticed that 
those positions do not depend on the pair-potential amplitude.
%

\subsubsection{Dependence of the inter-band hopping amplitude $t'$}
\begin{figure}
  \begin{center}
    \begin{tabular}{p{40mm}p{40mm}}
      \resizebox{55mm}{!}{\includegraphics{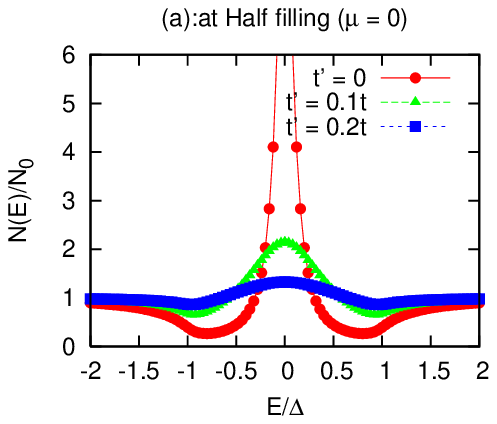}} &
      \resizebox{55mm}{!}{\includegraphics{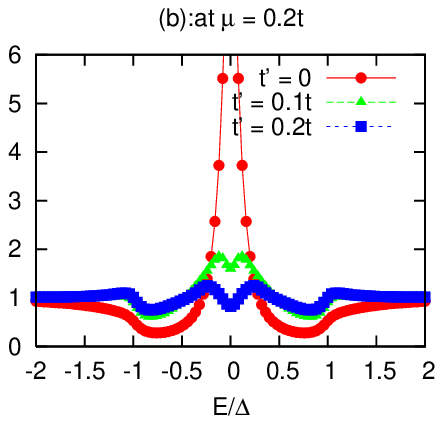}} 
    \end{tabular}
\caption{\label{fig:tdep}
(Color online)
The density of states at [110] surface for various inter-orbital hopping amplitude $t'$.
The pair potentials are  $\Delta_{A}  = \Delta = 0.001t$ and $\Delta_{B} = -\Delta_{A}$.
(a) the half filling ($\mu = 0$) and (b) $\mu = 0.2t$.  
The smearing factor is $\eta=0.1\Delta$.
}
  \end{center}
\end{figure}

We investigate the dependence on the inter-orbital hopping amplitude $t'$.
We consider [110] surface ($\theta = \pi/4$).
As shown in Fig.~\ref{fig:tdep}(a), the ZBCP always exists at the half filling ($\mu = 0$) for any inter-band hopping amplitudes $t'$. 
At $\mu = 0.2 t$ as shown in Fig.~\ref{fig:tdep}(b), the ZBCP only appears when without an inter-band hopping, i.e., $t'=0$.
These ZBCPs appear when the appearance condition in Eq.~(\ref{eq:anare}) is satisfied. 
%

\section{Discussion}
The advantages of our method are that 
one can easily investigate the surface-angle dependence of the density of states with the use of the quasiclassical method and 
easily calculate the density of states in the $n$-band (multi-band) system with less computational machine-time. 
Therefore, we can take, for example, a realistic 5 band model in order to discuss the density of states for iron-based superconductors. 
We will report its results elsewhere near future.
We have assumed that the matrix of the pair potential in the band-representation does not have off-diagonal elements, 
which correspond to the inter-band pairings. 
When the inter-band pairing is dominant, the Cooper pairs have center-of-mass momentum $\Vec{q} \neq 0$. 
Usually such pairs are not energetically favorable
since the pair potentials have spacial dependence even in bulk systems. 
Starting with the same Matsumoto-Shiba method,\cite{Matsumoto}
Onari {\it et al.}\cite{onari} recently calculated the surface Andreev bound states without the quasiclassical approximation.
Their results show that the peak positions of the Andreev bound states depend on the gap amplitudes
on two bands in the same two-band model as considered in Sec.~IV, 
and
the ZBCP does not always appear at the half-filling. 
These results might seemingly be inconsistent with our quasiclassical results.
It is, however, not the case.

They obtained the perturbed Green function
by directly integrating the original unperturbed Green function over $k_{x}$ and $k_{y}$ numerically. 
The original unperturbed Green function has sharp peaks on Fermi surfaces in the momentum space and 
rapid Fermi-wave-length oscillations in the real space.
We have integrated out those properties by the quasiclassical approximation.
It should be noted that the pair potentials are of the order $\Delta \sim 0.1t$ in Ref.~\onlinecite{onari}. 
 This parameter is out of our quasiclassical approach ($\Delta/t \ll 1$). 
 As shown in Sec.~IV.B.3,  
our analytical result, which depends on the details of the gap amplitudes
and therefore is consistent with Ref.~\onlinecite{onari},
is reduced to the quasiclassical result in the limit $\Delta/t \ll 1$.
 Thus, the differences in the obtained results between Onari {\it et al.}\cite{onari} and the present paper would be due to the difference in
applicable parameter regions.
The formulation derived in Secs.~II and III can be applied to general multi-band superconductors including $d$-wave pairing superconductor.
The appearance condition for the ZBCP
is given as Eq.~(\ref{eq:app}) in Sec.~III. 
Let us consider, for instance, the case of the two-band model discussed in Sec.~IV 
at the half filling for [110] surface. 
From Eq.~(\ref{eq:app}),
the appearance condition for the ZBCP is given as 
\begin{equation}
{\rm sgn}\: (\Delta_{A1}) +{\rm sgn}\: (\Delta_{A2}) +{\rm sgn}\: (\Delta_{B1}) +{\rm sgn}\: (\Delta_{B2}) =0.
\end{equation} 
Here, $\Delta_{A1}$ and $\Delta_{A2}$ are the pair potentials on the inner Fermi surface (red) in Fig.~\ref{fig:yokogiru}, and $\Delta_{B1}$ and $\Delta_{B2}$ are the pair potentials on the outer Fermi surface (green) there. 
For a two-band $d$-wave superconductor, $\Delta_{A1} = - \Delta_{A2}$ and $\Delta_{B1} = - \Delta_{B2}$,
so that the above condition is satisfied and the ZBCP appears.
Furthermore, in the case of a single-band $d$-wave superconductor, 
the appearance condition for the ZBCP is obtained from Eq.~(\ref{eq:app}) as 
\begin{equation}
{\rm sgn}\: (\Delta_{A1}) +{\rm sgn}\: (\Delta_{A2}) = 0.
\end{equation}
This is consistent with previous results for $d$-wave pairing in Refs.~\onlinecite{Hu,TanakaPRL,Kashiwaya,Matsumoto},
where the ZBCP appears when the quasiparticles feel a superconducting phase change $\pi$ in
the surface scattering process A1$\leftrightarrow$A2 on a Fermi surface.

\section{Conclusion}
In conclusion, we extended the single-band method by Matsumoto and Shiba\cite{Matsumoto} into general $n$-band case ($n \ge 2$).
With the use of the quasiclassical approximation, we developed the way to integrate the unperturbed Green function  with respect to $k_{x}$ which is the momentum component perpendicular to a surface. 
We showed that the appearance condition of the ZBCP does not depend on any anisotropy in the pair-potential amplitude, but only on the relative phase,
in the case of $\Delta \ll E_{F}$ in $n$-band systems. 
The properties of the normal state are influential for the ZBCP to appear. 

We also calculated the surface density of states in the two-band system as a simple example of our approach. 
We suggested that our quasiclassical approach is appropriate when $|\Delta|/t \ll 1$. 
We showed that the peaks of the density of states due to the Andreev bound states depend on the surface angle and 
the parameters in the normal state ($t$,$t'$, $\mu$),  %
so that the sign-reversing $s$-wave ($\pm s$-wave) superconductors exhibit complicate properties in the tunneling spectroscopy compared 
with single-band $d$-wave superconductors.

\section*{Acknowledgment}
We thank Y. Kato, M. Machida, N. Nakai, H. Nakamura, M. Okumura, C. Iniotakis, M. Sigrist, Y. Tanaka and  S. Onari for helpful discussions and comments.
Y.N.~acknowledges support 
   by Grand-in-Aid for JSPS Fellows (204840). 
N.H.\ is supported by JSPS Core-to-Core Program-Strategic Research Networks,
`Nanoscience and Engineering in Superconductivity (NES)'.

\appendix*
\section{Integration without quasiclassical approximation}
We will show the ZBCP appearance condition Eqs.~(\ref{eq:dab}) and (\ref{eq:iab})
for zero inter-orbital hopping 
amplitude ($t'=0$),
by integrating Eq.~(\ref{eq:g0wa}) in the 
simple two-band model for [110] surface at the half filling.\cite{note}

The Hamiltonian in the normal state is described as 
\begin{eqnarray}
\hat{H}^{o} = 
\left(\begin{array}{cc} 
\epsilon_{1}(\tilde{k}_{x},\tilde{k}_{y})  & 0 \\ 
0 & \epsilon_{2}(\tilde{k}_{x},\tilde{k}_{y}) 
 \end{array}\right),
\end{eqnarray}
with
\begin{eqnarray}
\epsilon_{1}(\tilde{k}_{x},\tilde{k}_{y}) &=& -t \cos(\tilde{k}_{x} - \tilde{k}_{y}), \\
\epsilon_{2}(\tilde{k}_{x},\tilde{k}_{y}) &=& -t \cos(\tilde{k}_{x} + \tilde{k}_{y}). 
\end{eqnarray}
Here, we have introduced $\tilde{k}_{x} = k_{x}/\sqrt{2}$ and $\tilde{k}_{y} = k_{y}/\sqrt{2}$. 
Considering the pair potentials $\Delta_{A,B}$ which do not depend on $\Vec{k}$ 
and using the unitary matrix Eq.~(\ref{eq:uni}),
the pair potential matrix in the orbital representation can be written as 
\begin{eqnarray}
\hat{\Delta}^{o} &=& \left(\begin{array}{cc} \Delta_{A} \theta(\tilde{k}_{x}) + \Delta_{B} \theta (-\tilde{k}_{x}) &  0\\
0&  \Delta_{A} \theta(\tilde{k}_{x}) + \Delta_{B} \theta (-\tilde{k}_{x})  \end{array}
\right), \nonumber \\
&\equiv&
\left(\begin{array}{cc} \Delta_{\tilde{k}_{x}} &  0\\
0&   \Delta_{\tilde{k}_{x}}  \end{array}
\right).
\end{eqnarray}
The unperturbed retarded Green function $\check{G}_{0}^{R}(E,k_{x},k_{y})$ is written as 
\begin{eqnarray}
\check{G}_{0}^{R}(E,\tilde{k}_{x},\tilde{k}_{y}) = (E - \check{H}_{\rm N}^{o})^{-1} 
&=&
\left(\begin{array}{cc} 
\hat{A}_{+}  & \hat{B} \\ 
\hat{B}& \hat{A}_{-}
 \end{array}\right),
\end{eqnarray}
where 
\begin{eqnarray}
\hat{A}_{\pm} &=& 
\left(\begin{array}{cc} 
\frac{E \pm \epsilon_1}{-|\Delta_{\tilde{k}_{x}}|^2 + E^2 - \epsilon_1^2} & 0 \\ 
0 & \frac{E \pm \epsilon_2}{-|\Delta_{\tilde{k}_{x}}|^2 + E^2 - \epsilon_2^2}
 \end{array}\right), \\
 \hat{B} &=& 
 \left(\begin{array}{cc} 
\frac{\Delta_{\tilde{k}_{x}}}{-|\Delta_{\tilde{k}_{x}}|^2 + E^2 - \epsilon_1^2} & 0 \\ 
0 & \frac{\Delta_{\tilde{k}_{x}}}{-|\Delta_{\tilde{k}_{x}}|^2 + E^2 - \epsilon_2^2}
 \end{array}\right).
\end{eqnarray}
To investigate the appearance condition of the ZBCP, we set $E = 0$ (i.e., zero energy),
$x = 0$ and $x' = 0$ (i.e., at the surface). 
Then, we calculate $\check{G}_{0}^{R}(E = 0,x=0,x'=0,k_{y})$:
\begin{eqnarray}
\check{G}_{0}^{R}(E = 0,x=x'=0,\tilde{k}_{y})
&=& 
\int_{-\pi}^{\pi}\frac{d \tilde{k}_{x}}{2 \sqrt{2} \pi}  \check{G}_{0}^{R}(E=0,\tilde{k}_{x},\tilde{k}_{y}). \nonumber \\
\end{eqnarray}
Each element in this matrix can be integrated analytically as
\begin{eqnarray}
\int \frac{ d \tilde{k}_{x} \cos(\tilde{k}_{x} \pm \tilde{k}_{y})}{|\Delta|^2  + \cos^2(\tilde{k}_{x} \pm \tilde{k}_{y})}
&=&  \int  \frac{ - dx}{|\Delta|^2  + 1- x^{2}}, 
\\
\int  \frac{d \tilde{k}_{x}}{|\Delta|^2  + \cos^2(\tilde{k}_{x} \pm \tilde{k}_{y})}
&=& \int  \frac{dx \frac{1}{1+x^{2}}
}{|\Delta|^2  +\frac{1}{1+x^{2}}}. \: \: \: \: \: \: \: \: \: \: \: \: 
\end{eqnarray}
Integrating $\check{G}_{0}^{R}(E=0,\tilde{k}_{x},\tilde{k}_{y})$
by using the above formulae, we finally obtain 
$\check{G}_{0}^{R}(E = 0,x=0,x'=0,k_{y})$: 
\begin{eqnarray}
\check{G}_{0}^{R}(E = 0,x=0,x'=0,k_{y}) \propto \frac{1}{t}
\left(\begin{array}{cccc}
-I_{1} & 0 &\Delta_{ab}
  & 0 \\
0 &-I_{2}& 0 &\Delta_{ab}  \\
\Delta_{ab} & 0 & I_{1} & 0\\
0  & \Delta_{ab} & 0 & I_{2}
\end{array}\right), \nonumber \\
\end{eqnarray}
where $I_{1,2}$ and $\Delta_{ab}$ are defined
in Eqs.~(\ref{eq:qui}) and (\ref{eq:qud}).
Its inverse matrix is written as 
\begin{eqnarray}
[\check{G}^{R }_{0}]^{-1} &\propto& t 
\left(\begin{array}{cccc}
\frac{-I_{1}}{(\Delta_{ab})^2  + I_{1}^2} & 0 & \frac{ \Delta_{ab}}{(\Delta_{ab})^2  + I_{1}^2} & 0 \\
0 &\frac{-I_{2}}{(\Delta_{ab})^{2} +  I_{2}^{2}} & 0 & \frac{\Delta_{ab}}{(\Delta_{ab})^2 + I_{2}^2} \\
 \frac{ \Delta_{ab}}{(\Delta_{ab})^2  + I_{2}^2} & 0 & \frac{I_{1}}{(\Delta_{ab})^2  + I_{1}^2} & 0 \\
0  & \frac{\Delta_{ab}}{(\Delta_{ab})^2 + I_{2}^2} & 0 &\frac{I_{2}}{(\Delta_{ab})^{2} +  I_{1}^{2}} 
\end{array}\right). \nonumber \\
\end{eqnarray} 
The zero energy bound states appear when
$[\check{G}^{R }_{0}]^{-1}$
diverges as noticed from Eqs.~(\ref{eq:4}) and (\ref{eq:pg}).
Therefore, the appearance condition of the ZBCP is expressed as 
\begin{eqnarray}
\Delta_{ab} &=& 0, \\
I_{1} = 0 \: \: &{\rm or}& \: \: I_{2} = 0.
\end{eqnarray}

\end{document}